\documentstyle[12pt]{article}
\title{Black holes radiate but do not evaporate}
\author{Hrvoje Nikoli\'c \\
Theoretical Physics Division, Rudjer Bo\v{s}kovi\'{c} Institute, \\
P.O.B. 180, HR-10002 Zagreb, Croatia \\
{\normalsize hrvoje@thphys.irb.hr} \\
\makebox[1in]{} \\
}
\date{\today}
\begin{document}
\maketitle
\begin{abstract}
During the black hole radiation, 
the interior contains
{\em all} the matter of the initial black hole, together with  
the negative energy quanta entangled with the exterior Hawking 
radiation. Neither the initial matter nor the negative energy quanta
evaporate from the black hole interior.
Therefore, the information is not lost during the radiation. 
The black hole mass eventually drops to zero
in semiclassical gravity, but this semiclassical state
has an infinite temperature and still 
contains all the initial matter together with the negative energy 
entangled with the exterior radiation.
\end{abstract}
\vspace*{0.5cm}
PACS Numbers: 04.70.Dy \newline
{\it Keywords}: Hawking radiation; black hole information
\vspace*{0.9cm}

The semiclassical theory of quantum fields in gravitational backgrounds
suggests that black holes evaporate by
radiating particles. The spectrum of radiated particles is
thermal \cite{hawk,bd}. At large distances from the black hole,
the temperature (in units $\hbar=c=G_{\rm N}=k_{\rm B}=1$) is equal to
\begin{equation}\label{temp}
T=\frac{1}{8\pi M},
\end{equation}
where $M$ is the mass of the black hole. However, a thermal state
is a mixed state. This implies that the final state after the 
complete black hole (BH) evaporation is a mixed state, 
even if the initial state was pure. The evolution of a pure state 
into a mixed state
is in contradiction with the principle of unitary evolution,
which constitutes the notorious BH information paradox.

In order to solve the BH information paradox, various 
refinements of the original calculation \cite{hawk} 
have been proposed (for reviews, see, e.g., 
\cite{har,pag,gid,str}), but none 
of the proposals is considered as the definite solution 
of the problem. Typical refinements include arguments 
that the radiation is not exactly thermal so that the 
information is contained in long-range correlations among 
radiated particles, or that the Planck-scale physics modifies 
the semiclassical approximation so that the final state of 
the black hole is a stable (or long living) remnant which has the 
mass and radius on the Planck scale and contains all the information. 
Although some of these refinements are probably more realistic 
then the original semiclassical calculation \cite{hawk}, 
in this essay we argue that none of these refinements is 
essential for solving the BH information paradox. Instead, 
the semiclassical calculation with assumption of the exact 
thermal spectrum, although probably not fully realistic, 
seems to be sufficient for solving the 
BH information paradox itself.   

We start with the observation that whenever the process of
particle creation is described by a Bogoliubov transformation of
the initial pure state, the final state is also a pure (squeezed)
state \cite{gris}. The description of such a state in terms
of a mixed state is only an artefact of the tracing out of the
unobserved degrees of freedom. In the case of a black hole, one
traces out the degrees of freedom that correspond to the
BH interior hidden by the horizon, which leads to a
mixed thermal density matrix that describes the BH exterior
\cite{gris,bd}. (To be clear, the horizon is also considered 
as a part of the BH interior.)
At first sight, this observation seems to solve 
the BH information paradox, because the information is 
contained in the quantum correlations between the exterior and 
interior degrees of freedom. However, this only postpones 
the problem to the moment when the black hole evaporates 
completely, so that at the end there is no a BH interior for 
the Hawking radiation to be correlated with. This lack of the
BH interior in the final state is 
the reason for the assumption that the black hole does not evaporate 
completely, but ends in a Planck-scale remnant. 
In this essay, we argue that the assumption 
of a Planck-scale remnant is unnecessary for 
the solution of the BH information paradox. Instead, even if we 
assume that the black hole radiates thermal radiation 
until the BH mass drops to zero, there is still an object with 
which the radiation is correlated. In fact,  
Eq.~(\ref{temp}) suggests that this must be so because it says that 
if the final state of the black hole has zero mass, then its 
temperature is {\em infinite}. The infinite temperature 
suggests that the final state is not simply a vacuum state. 
This claim can be made more precise in the following way. 
Consider the BH metric
\begin{equation}
ds^2=\left( 1-\frac{2M}{r} \right) dt^2 -
\displaystyle\frac{1}{1-\displaystyle\frac{2M}{r}} dr^2 -r^2d\Omega^2 .
\end{equation}
When $M=0$, then, at first sight, this metric seems to correspond 
to the flat metric $ds^2=dt^2 -dr^2 -r^2d\Omega^2$. However, this is 
not really true for $r=0$. The case $r=0$ requires a more careful 
calculation. In particular, we have
\begin{equation}
\lim_{M\rightarrow 0^+} g_{00}(r=2M)=0, \;\;\;\;
\lim_{M\rightarrow 0^+} g_{rr}(r=2M)=\infty .
\end{equation}
This limit corresponds to the existence of the horizon at $r=0$ 
for $M=0$, which is the origin of the infinite temperature 
for $M=0$. This is, indeed, what one expects physically, 
because if a horizon exists for an arbitrarily small $M$ 
(which must be the case if we assume that the radiation is thermal 
even for small $M$'s), then, in the sense of a limit, the horizon 
must also exist for $M=0$. Thus, when $M$ drops to zero, a 
metric singularity remains at $r=0$. 

The discussion above shows that a geometrical defect remains 
when the BH mass drops to zero. To explore the 
matter content of this defect, note first 
that the Hawking radiation is determined purely by the geometry 
of spacetime and does not rest on the validity of the Einstein 
equations or on the matter content of the source for the 
gravitational field \cite{viss1,viss2}. In fact, the Hawking radiation
from a collapsing black hole 
is just a special case of the general result that a time-dependent
gravitational background leads to particle creation \cite{bd}. 
An instructive example is the Milne spacetime \cite{bd}, which is a 
time-dependent flat spacetime {\em without matter}. The Milne 
spacetime also creates particles from the vacuum. Obviously, 
the conservation of energy during particle creation cannot be 
related to a loss of energy of some previously existing matter.
Instead, the creation of positive-energy 
quanta is allways accompained with the creation of 
negative-energy quanta, so that the total energy is conserved.
In a more precise language based on the notion of local renormalised 
energy-momentum tensor, there is a flux of negative energy that 
falls into the black hole \cite{bd}. 
(For other details on the theory of negative energies in 
quantum field theory see also \cite{bro,for1,for2,for3}.)
Therefore, the Hawking radiation is not associated with 
a decrease of mass of the {\em initial} BH matter or with an 
escape of that matter from the black hole.
(By the initial BH matter
we understand the matter of which the black hole was built 
just before the Hawking-radiation 
process has started.) All the 
information contained in that matter remains in the interior. 
Similarly, the negative energy quanta entangled and correlated 
with the outgoing Hawking radiation fall into the black hole 
and remain there. The contribution of the negative energy quanta 
to the total BH mass cancels with that of the initial matter,
but it does not mean that their informations cancel. Thus, the 
information does not need to possess a mass or energy. 
(The usual argument that information requires energy 
\cite{bek1,bek2} does not apply here owing to the existence of negative energy 
that also may carry information.)
Therefore, nothing 
escapes from the black hole during the BH 
radiation. Neither matter, nor information.
In particular, the Hawking radiation 
does not escape {\em from} the black hole simply because 
it is created near but still outside of the horizon. 
(If it was created at the horizon itself, then it would need an infinite
time to escape. If it was created far from the horizon, then its 
spectrum would not be thermal.) In other words, {\em 
black holes produce Hawking radiation, but nothing evaporates from 
black holes}. This, together with the fact that the negative energy
quanta in the interior are entangled with the Hawking radiation,
constitutes our solution of the BH information paradox!
Thus the semiclassical BH radiation does 
not contradict the classical law that nothing can escape from a 
black hole. Note also that
this solution does not contradict the classical 
theorem that black holes ``do not have hair", because the 
nonlocal information 
contained in the entanglement between the interior and exterior 
does not change the fact that the local exterior metric depends only 
on the total mass contained in the interior.
 
Owing to the absorption of negative energy,
semiclassical gravity predicts that  
the BH mass eventually drops to zero.
One may be worried by the fact that this massless 
pointlike object with the vanishing surface may contain 
an arbitrary amount of information. Moreover, a part of this information, 
i.e., the information contained in the initial BH matter, is 
of purely local origin. Indeed, this property may be viewed 
as indicating a pathology of semiclassical theory, just as 
classical singularity theorems indicate a pathology of 
classical general relativity. In fact, obtaining such a pathology 
should not be surprising, because, in field theory with a 
background metric, an arbitrarily small volume contains an 
infinite number of points, and thus an infinite number of the 
degrees of freedom that may store the information.
It is very likely that the full 
quantum gravity fixes this pathology. 
However, this pathology does 
not directly influence our main conclusion that the information is 
stored in the BH interior and in the correlations 
of the interior with the exterior. In particular, as the apparent 
horizon is the only essential geometrical ingredient for Hawking 
radiation \cite{viss2}, the existence of a singularity in 
a classical or a semiclassical description of a 
black hole is not a serious argument against this conclusion. 
As a mathematical description of a system with a singularity 
is ambiguous, one can introduce an appropriate  
regularization of the singularity by hand, thus avoiding
the singularity on a purely technical level. The main conclusion 
above does not depend on the exact way the singularity has been 
regularized.

Finally, let us also note that our discussion solves an often 
mentioned problem \cite{har,gid,str} with the light remnant scenario, 
consisting in the claim that light remnants containing a huge information 
should be often produced in various physical processes, which is 
not seen in nature. According to our discussion, the remnant is light owing
to the negative energy in it. However, this negative energy cannot 
exist without the exterior Hawking radiation with which it is 
entangled. Therefore, the total system containing 
a huge information -- the light remnant entangled
with the Hawking radiation -- is not light at all, so one does not expect 
such systems to be often produced in various physical processes.

To conclude, we stress that
our semiclassical analysis cannot 
offer a realistic solution to the problem of the final state, 
so it would be incorrect to say that our analysis completely 
solves the BH information {\em problem}. Nevertheless, 
our analysis indicates that, at least, the semiclassical 
treatment does not lead to an information {\em paradox}. 
In other words, the semiclassical analysis is not completely 
realistic, but, at least, seems to be self-consistent in 
not contradicting the unitarity of the semiclassical theory.
Even if the prediction of a singularity is viewed as an 
internal inconsistency of the semiclassical theory,
our analysis indicates that this inconsistency does {\em not}
necessarily lead to the violation of unitarity. 
If information is conserved at {\em each} moment of time before 
the final singularity, then, in the sense of a limit, it is 
also conserved at the time at which the singularity begins.

\section*{Acknowledgments}

This work was supported by the Ministry of Science and Technology of the
Republic of Croatia under Contract No.~0098002.

\end{document}